# Surface Wave-Based Underwater Radio Communication

Igor I. Smolyaninov, Quirino Balzano, Christopher C. Davis, and Dendy Young

*Abstract*—An underwater portable radio antenna operating in the 50 MHz band and efficient for launching surface electromagnetic waves at the seawater/air interface is presented. The antenna operation is based on the field enhancement at the antenna tip and on an impedance matching antenna enclosure, which is filled with de-ionized water. This enclosure allows us to reduce antenna dimensions and improve the coupling of electromagnetic energy to the surrounding salt water medium. Since surface wave propagation length far exceeds the skin depth of conventional radio waves at the same frequency, this technique is useful for broadband underwater wireless communication over several meters distances.

## I. Introduction

It is very difficult to employ radio signals for long distance undersea communication because they are rapidly attenuated in fresh, brackish, and salt water. Performance of conventional RF communication schemes in water is limited by the relatively small RF skin depth, which may be estimated as:

$$\delta = \sqrt{\frac{1}{\pi\mu_0\sigma\nu}} \approx \frac{270 Hz^{1/2} m}{\sqrt{\nu}} \qquad (1)$$

where $\sigma$ is the water conductivity and $\nu$ is the communication frequency [1]. This signal attenuation severely limits the ability to communicate over distance in water. For example, at $\nu$ = 50 MHz the skin depth in seawater is about 3.8 cm, so conventional techniques of RF communication are impractical in salt water over useful distances.

On the other hand, efficient coupling to surface electromagnetic modes [2] (such as the surface electromagnetic wave at the seawater/air interface) may enable much longer communication distances. The penetration depth $L_z$ and the propagation distance $L_r$ of a surface wave are given by the following expressions:

$$L_z \approx \frac{\lambda_0}{4\pi\sqrt{\varepsilon''}}, \qquad (2)$$

and

$$L_r \approx \frac{\lambda_0 \varepsilon''}{\pi}, \qquad (3)$$

respectively [3], where $\varepsilon''$ is the imaginary part of the dielectric constant of seawater and $\lambda_0$ is the free space wavelength. Thus, at 50 MHz the signal propagation distance appears to be rather large ($L_r$ = 60 m), while (assuming operation down to −90dB relative signal levels) the communication depth may reach several meters. This argument is illustrated in Fig.1, which shows the RF field distribution in seawater produced by a point source located in the vicinity of the seawater/air interface. The term "seawater" will be used throughout this letter to describe a water environment with approximately 3.5% salinity and a dielectric constant of ~81. At some distance from a source the field distribution is dominated by the surface electromagnetic wave, which enables RF communication from point A to point B, which would otherwise be impossible in the absence of the surface wave due to the skin effect in bulk seawater.

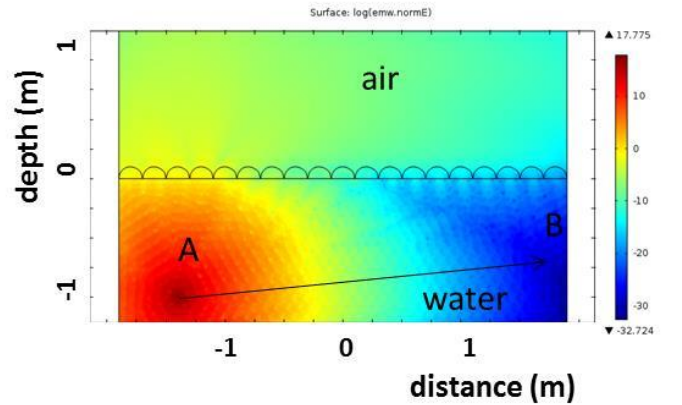

Fig. 1. Numerical simulations of RF field distribution in seawater (shown in logarithmic scale) produced by a 50 MHz point source located in the vicinity of the seawater/air interface performed using the COMSOL Multiphysics solver. At some distance from a source the field distribution is dominated by the surface electromagnetic wave. Note that the signal propagating just below the wavy water surface is stronger than the signals propagating in the bulk water.

Below we will demonstrate that an underwater radio antenna operating in the 50 MHz band, suitable for efficient excitation of surface electromagnetic waves at the seawater/air interface can be realized. Since surface wave propagation length far exceeds the skin depth of conventional radio waves at the same frequency, such an antenna may be useful for broadband underwater wireless communication over several meters distances.

## II. Antenna Design and Discussion

There is a rather extensive literature on the propagation of radio waves in sea water [4] and on the performance of antennas in such environment. Siegel and King [5] demonstrated that practical size antennas can be modeled as electrically short dipoles and exhibit dipole behavior in sea water. Yoshida [6] points out that it is necessary to keep the entire communication system below water to avoid above water transmitter-receiver coupling. In [7], the authors present a detailed analysis of sea rough upper surface effects on radiated fields. In this work our antenna design followed the analysis developed in [8] to evaluate the RF fields radiated from a short dipole in a highly dissipative environment, human tissue. In [8] a

I. I. Smolyaninov is with the Saltenna LLC, 1751 Pinnacle Drive, Suite 600 McLean VA 22102-4903 USA (phone: 443-474-1676; e-mail: igor.smolyaninov@saltenna.com).

Q. Balzano is with the Electrical and Computer Engineering Department, University of Maryland, College Park, MD 20742 USA (e-mail: qbalzano@umd.edu).

C. C. Davis is with the Electrical and Computer Engineering Department, University of Maryland, College Park, MD 20742 USA (e-mail: davis@umd.edu).

D. Young is with the Saltenna LLC, 1751 Pinnacle Drive, Suite 600 McLean VA 22102-4903 USA (e-mail: dendy.young@saltenna.com).

short dipole is enclosed in an insulating layer to avoid the RF losses caused by the high E- fields at the tips of an antenna immersed in a highly dissipative medium.

In this paper, following the analysis in [8], a helical dipole resonant at 50 MHz, which is more efficient than a linear short of the same length, has been used as a radiating structure, enclosed by a substantial insulating layer of deionized water to minimize reactive near field losses and to provide a minimal matching structure for wave propagation in sea water. A photograph of the 50 MHz underwater antennas optimized for surface electromagnetic wave excitation is shown in Fig. 2. The antenna operation is based on the field enhancement at the antenna tip and on the impedance matching antenna enclosure, which is filled with de-ionized water. This enclosure allows us to reduce antenna dimensions by approximately factor of 9 compared to free space dimensions, and to improve coupling of electromagnetic energy to the surrounding seawater medium [9]. In addition, it considerably reduces the ohmic losses that would be caused by the immersion of the antenna in saline water.

The impedance matching enclosure around the antenna uses an impedance matching fluid. De-ionized water can be such a fluid, since it has about the same dielectric constant as seawater and much smaller conductivity. As illustrated in Fig.3, laboratory testing of the de-ionized water based impedance-matching enclosure concept performed in brackish water (0.5% salinity) at 2.45 GHz indeed indicated the considerable advantage of such an enclosure over a conventional antenna design. However, we also mention that other impedance matching fluids and/or media may be engineered for the same purpose of drastically reducing the near-field antenna losses [9].

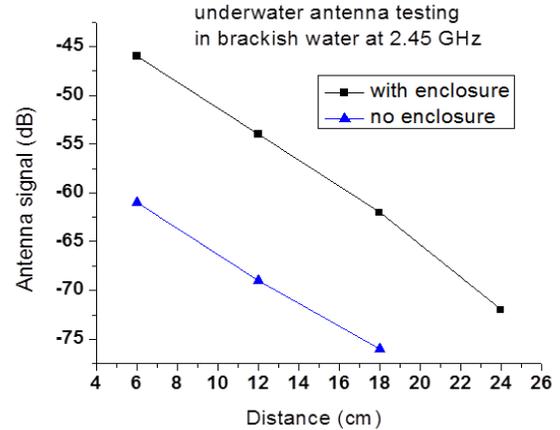

Fig. 3. Laboratory testing of the impedance-matching enclosure concept performed in brackish water (0.5% salinity) at 2.45 GHz. The transmitter-receiver set equipped with impedance-matching enclosures shows a ~20 dB advantage compared to the antennas directly immersed in brackish water. The comparative advantage grows quickly with increased salinity, since unenclosed operation of these antennas in seawater becomes impossible due to losses.

The radiating structure selected to maximize the electric field at its tip was a helical monopole (shown in Fig.4). The length and the diameter of the helical antenna were selected for resonance in de-ionized water at 50 MHz. The monopole helix, with a ground plane, resonant at 450 MHz was immersed in a water bath and trimmed for resonance at 50 MHz. The tuning was further refined by immersing the antennas in a gallon container of deionized water and using a network analyzer (see Figures 5 and 6). In addition, the tapping point for a 50 Ω match to a feeding coaxial line was also determined by experimental tests using the setup of Fig. 5. The finalized helical monopole was 16 cm long with 0.7 cm diameter. The electric field at the antenna tip was further maximized by tip sharpening.

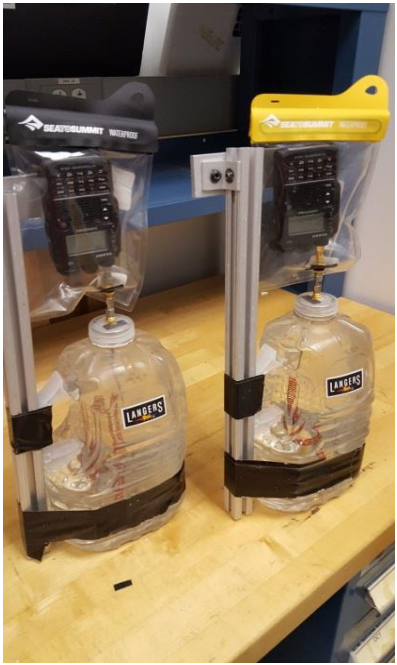

Fig. 2. Surface wave underwater antennas attached to Yaesu VX-8 radios operated at 50 MHz at 5W output power. The impedance-matching enclosures are filled with de-ionized water.

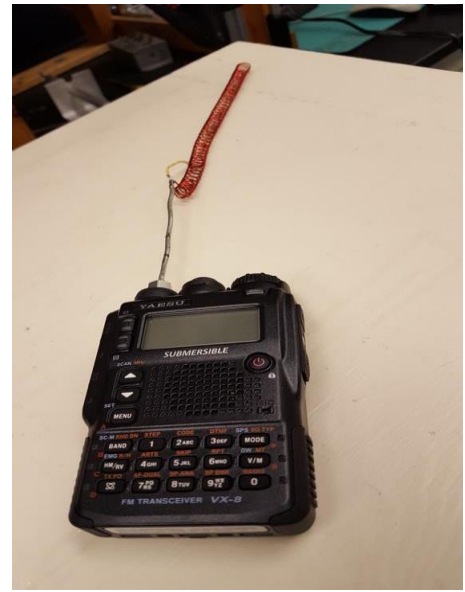

Fig. 4. Photo of the 50 MHz helical monopole antenna without the impedance-matching enclosure. Also shown is a Yaesu VX-8 radio.

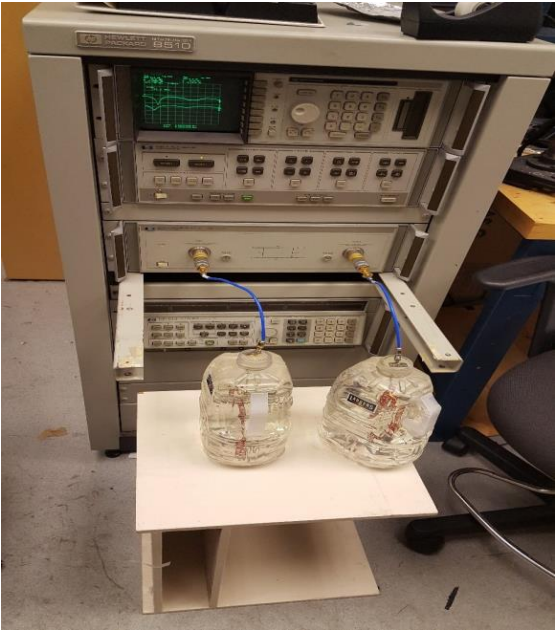

Figure 5. Preliminary tuning and matching of helical antennas.

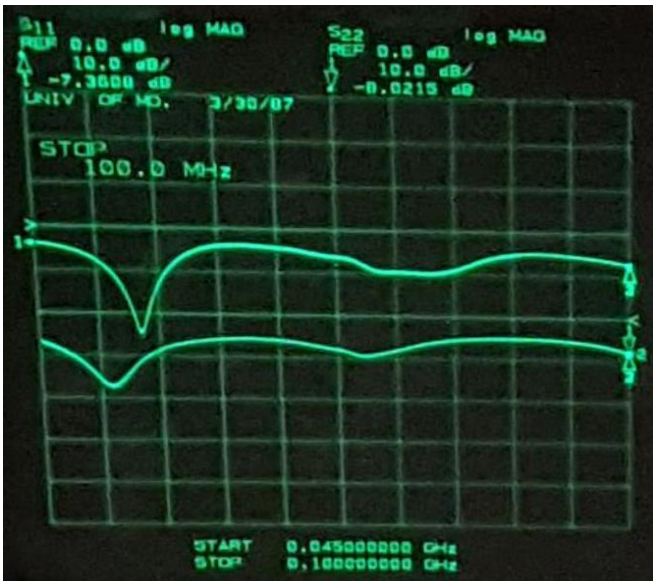

Figure 6. Measured $S_{11}$ and $S_{22}$ of the helical antennas in the configuration shown in Fig.5.

Two Yeasu submersible water proof 5 W transceivers, model VX-8DR/DE, were used to test the antenna performance. The helix structure was fed with a coaxial line 11 cm long. One end of the helix was short circuited to the coaxial ground (also the ground of the transceiver), while the center conductor of the coax line was used to feed the metal helix at 11 turns away from the ground point. Fig. 4 shows the assembled helical monopole and transceiver (without the impedance-matching enclosure). The final tuning of the antenna was performed by maximizing the received signal at 0.5 m distance using the two identical structures shown in Fig. 2.

III. MEASURED RESULTS

The operational performance of the surface wave antennas described above has been tested in an underwater, saline environment near Panama City, Florida (water salinity 2.8-3.2% depending on time of the day). The operating frequency was 50 MHz. The seawater environment was large enough (water depth at least 3 m) that no significant boundary effects were present. The effects of boundaries were tested separately by dedicated measurement runs performed near the sea floor and near the water basin borders. These boundaries were determined to have no effect on the measurement results. All the tests were conducted with separate transmitting (TX) and receiving (RX) antenna and radio systems enclosed in watertight containers (shown in Fig.2) which were handled by divers. The divers verified their depth using fixed vertical markers. The distance between divers was measured by using fixed horizontal markers. It was verified that the markers (made of buoys and ropes – see Fig.7) located inside the seawater had no effect on signal propagation between the radios. Signal propagation data read from the LED indicator and the S-meter of the Yaesu radios were reported by the divers and recorded by test personnel located on a nearby vessel. A picture of the Panama City sea water test conditions is presented in Fig. 8 showing two divers and their surface buoys

The communication range tests were conducted using the following steps at each depth/distance combination:

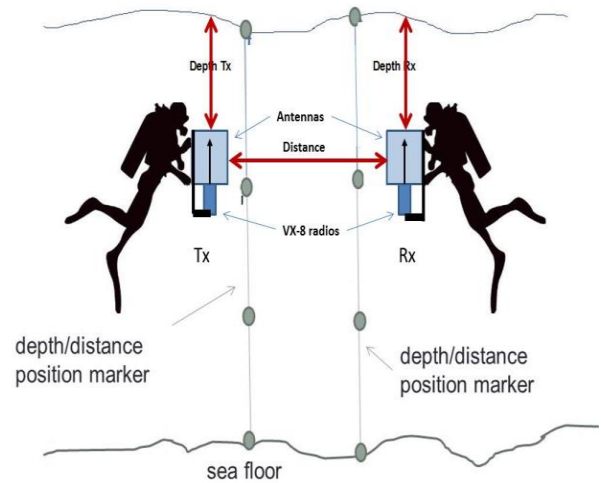

Figure 7. Schematic of underwater signal propagation range measurements

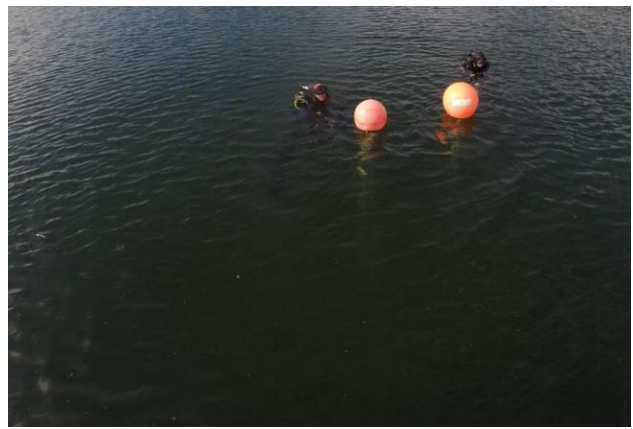

Fig. 8. Photo of the surface of the actual under water test range.

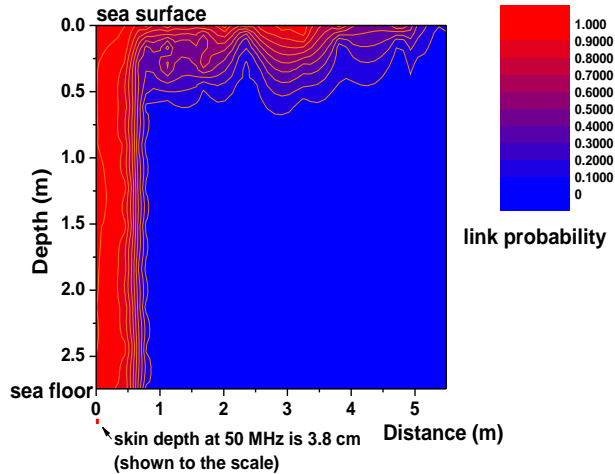

Fig. 9. Contour plot of the link probability as a function of depth/distance combination measured in seawater. The sea floor in these experiments was located at 3 m depth. The skin depth in seawater at 50 MHz is shown to the scale for comparison.

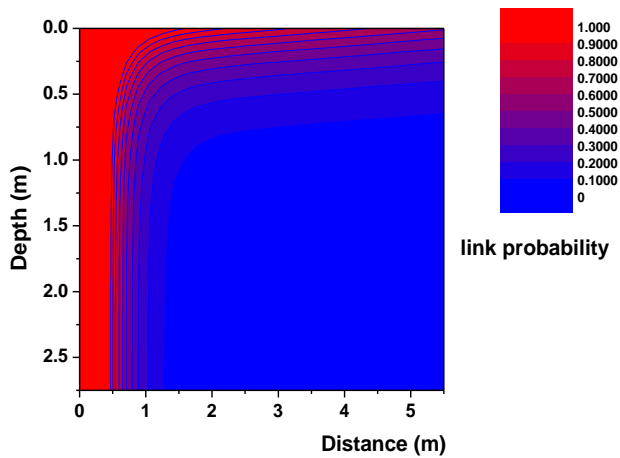

Fig. 10. Theoretical modeling of experimental plot in Fig. 9 assuming $L_r$=9 m value for the propagation distance of the surface electromagnetic wave.

1) Two divers holding the TX and RX units, respectively positioned themselves in seawater at the depth/distance markers.

2) The first diver pushed the PTT button on the TX unit 10 times for 1 s with 1 s intervals.

3) The second diver observed and reported the green LED RF link indicator and the signal level meter of his RX unit and reported if the RF link between the units has been established at the given depth/distance combination.

The averaged measured values of link probability are plotted in Fig. 9. The skin depth at 50 MHz in seawater is 3.8 cm. It is shown near the bottom left corner of the plot for comparison. At large depth the measured communication distance is limited to less than 1 m because of the skin effect in the bulk salt water. However, Fig. 9 clearly demonstrates that near the surface the novel 50 MHz underwater surface wave launching antennas described above enable RF communication range/depth combinations which far exceed the known 3.8 cm skin depth of seawater at 50 MHz. Relatively large variations of the link probability shown in Fig. 9 may be plausibly attributed to the variations in the sea surface wave ripples and salinity during the measurements. For comparison, Fig. 10 shows results of theoretical modeling of the experimental plot in Fig. 9 assuming $L_r$=9 m value for the propagation distance of the surface electromagnetic wave. The difference between this value and the 60 m theoretical estimate given by Eq.(3) may be attributed to the time-varying wavy surface of sea water, which has not been considered in [3]. Deteriorating effect of interface roughness on the propagation distance of surface electromagnetic wave is well established in the literature [2]. Even though smaller than the theoretical limit, the possible applications of such underwater link may include short-range diver-to-diver and underwater vehicle communication.

IV. CONCLUSION

Regardless of the relatively strong signal variations mentioned above, the antenna performance during the underwater tests generally matched theoretical expectations. According to our theoretical simulations presented in Fig. 1, at large distances from the source, propagation of radio waves beneath undulated sea water surface is dominated by surface electromagnetic waves. Experimental results plotted in Fig. 9 strongly support this theoretical prediction. We expect that further optimization of the antenna parameters will result in approaching the theoretical performance depth/distance limits described by Eqs. (2,3), which will enable novel capacity for wide bandwidth RF signal transmission through seawater, while reducing the power requirements, weight and size of the transmitting and receiving units.

2